\documentclass[fleqn,usenatbib]{mnras}

\usepackage{newtxtext,newtxmath}

\usepackage[T1]{fontenc}
\usepackage{ae,aecompl}

\usepackage{graphicx}   
\usepackage{amsmath}    
\usepackage{amssymb}    

\usepackage{breakurl}

\newcommand{\kms}{~km s$^{-1}$~}
\newcommand{\WR}{WR~146~}
\newcommand{\WRE}{WR~146}
\newcommand{\dotM}{~M$_{\odot}$~yr$^{-1}$~}
\newcommand{\XMM}{{\it XMM-Newton~}}

\newcommand{\Chandra}{{\it Chandra~}}
\newcommand{\ChandraE}{{\it Chandra}}
\newcommand{\Rosat}{{\it ROSAT~}}
\newcommand{\RosatE}{{\it ROSAT}}
\newcommand{\xspec}{{\sc xspec~}}
\newcommand{\xspecE}{{\sc xspec}}

\title[X-rays from \WRE]
{X-rays from the colliding wind binary \WR }

\author[S.A.Zhekov]{Svetozar A. Zhekov\thanks{
E-mail: szhekov@astro.bas.bg} \\
Institute of Astronomy and National Astronomical Observatory
(Bulgarian Academy of Sciences),\\
72 Tsarigradsko Chaussee  Blvd., Sofia 1784, Bulgaria
}

\date{}

\pubyear{2017}

\begin{document}
\label{firstpage}
\pagerange{\pageref{firstpage}--\pageref{lastpage}}
\maketitle

\begin{abstract}
The X-ray emission from the massive binary \WR is analysed in the 
framework of the colliding stellar wind (CSW) picture. 
The theoretical CSW model spectra match well the shape of the observed
X-ray spectrum of \WR but they overestimate considerably the observed
X-ray flux (emission measure). 
This is valid both in the case of 
complete temperature equalization and in the case of
partial electron heating at the shock fronts (different electron and 
ion temperatures), but, 
there are indications for 
a better correspondence between model 
predictions and observations for the latter. 
To reconcile the model 
predictions and observations, the mass-loss rate of \WR must be 
reduced by a factor of 8 - 10 compared to the currently accepted 
value for this object
(the latter already takes clumping into account). 
No excess X-ray absorption is derived from the
CSW modelling.
\end{abstract}




\section{Introduction}
\label{sec:intro}
Massive stars of early spectral types, OB and Wolf-Rayet (WR) stars,
posses fast and massive stellar winds 
(V$_{wind} = 1000-5000$\kms; 
$\dot{M}$ $\sim$ 10$^{-7} -  10^{-5}$ \dotM)
that play an important role for their evolution. 
On the other hand, when a binary harbours two massive stellar 
components, the interaction of their massive winds gives rise to the 
phenomenon of colliding stellar winds (CSW), whose observational
manifestation allows us to
study the physics of strong shocks in some detail. Due to the high
velocity of the stellar winds from massive stars, the shocked plasma
in the CSW region is heated to high temperatures, thus, it is
expected to be a strong source of X-ray emission 
(e.g., \citealt{pri_us_76}; \citealt{cherep_76}).

We note that the first systematic X-ray survey of WRs with
the {\it Einstein} Observatory showed that  WR$+$O
binaries are the brightest X-ray sources amongst them \citep{po_87},
but the quality of the X-ray data was not very high. 
With the launch of the \Chandra and \XMM observatories the number of
WR$+$O binaries with good quality X-ray spectra has increased
and some of the objects were studied in considerable detail.
Thus, in the last two-three decades it became generally accepted that
the enhanced X-ray emission from 
WR$+$O binaries likely originates 
from the interaction region of the winds of the massive binary 
components (see \citealt{rauw_naze_16} for a recent review on the
progress of studies, both theoretical and observational, of X-ray 
emission from interacting wind massive binaries
of early spectral type).

However, it is our understanding that carrying out detailed comparison 
between theory and observations is the most reliable way to test the 
currently accepted physical picture of a given phenomenon. 
In this respect, the modern tools for analysing and modelling of
observational data allow us to perform {\it direct} confrontation of
our physical ideas (models) with observations, and it is indeed the
case in the X-ray astronomy.

This is exactly the goal of the current study of the colliding stellar
wind phenomenon, namely, to carry out a direct comparison of the CSW
model results and the X-ray observations of the massive Wolf-Rayet 
binary \WRE.
Previous studies based on such an approach provided us with valuable
pieces of information on the physical picture of CSW binaries (e.g.,
\citealt{zhsk_00}; \citealt{zhp_10b}; \citealt{zh_15}). 
We believe that the more objects are analysed in such a way the better 
our understanding of the CSW phenomenon is.

Our paper is organized as follows.
We summarize  information on  \WR
in Section~\ref{sec:wrstar}. In Section~\ref{sec:observations}, we
review the recent X-ray observations of \WRE. In
Section~\ref{sec:csw_spec}, 
we present results from a {\it direct} comparison of the CSW model
with the X-ray emission of \WRE. In
Section~\ref{sec:discuss}, we discuss our results, and we present our
conclusions in Section~\ref{sec:conclusions}.

\begin{figure*}
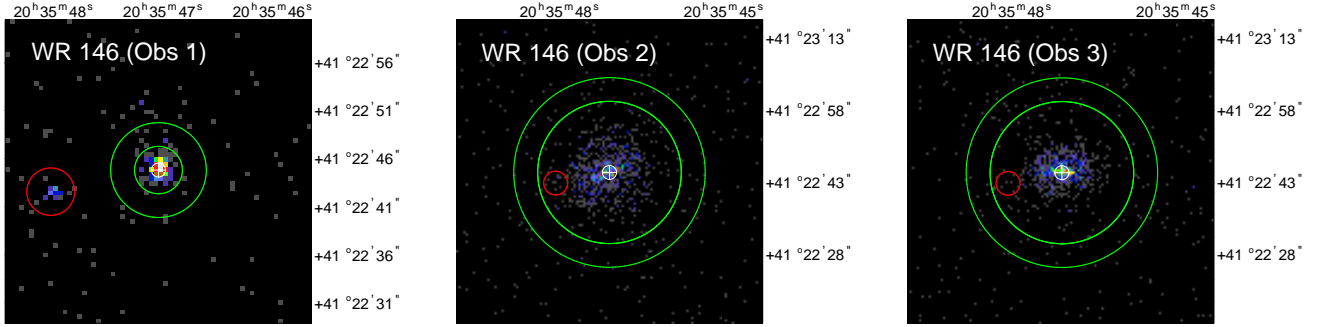

\includegraphics[width=2.30in, height=1.87in]{fig1a.eps}
\includegraphics[width=2.30in, height=1.87in]{fig1b.eps}
\includegraphics[width=2.30in, height=1.87in]{fig1c.eps}
\caption{The raw ACIS-I images of \WR in the 0.3 - 10 keV energy
band with the spectral extraction regions. The source spectrum was
extracted from the central circle, while the background spectrum was
extracted from adjacent annulus. The circled plus sign gives the
optical position of \WR (SIMBAD). 
For Obs 1, position of the near-by source BD$+40 4243$ is 
marked by a red circle. 
The source contribution to the spectrum of \WR was excluded for Obs 2 
and Obs 3 (again noted by a red circle).
}
\label{fig:image}
\end{figure*}

\section{The Wolf-Rayet Star \WR}
\label{sec:wrstar}
\WR (MR 112) is a massive binary whose components are a carbon-rich (WC) 
Wolf-Rayet star and an O-type star \citep{vdh_01}. 
It is a {\it wide}
colliding wind binary showing non-thermal radio emission
\citep{doug_96}. High-resolution radio observations resolved its
emission into three different sources: two thermal sources (identified
with the WR and O star, respectively) and one non-thermal source 
(associated with the CSW region in the binary) with an estimated binary 
separation of $0\farcs162\pm0\farcs008$ \citep{doug_00}. Similar result 
for the binary separation ($0\farcs168\pm0\farcs031$) was obtained
from the optical observations with the {\it Hubble Space Telescope 
(HST)} \citep{niemela_98}.
It is interetsing to note that \citet{setia_00} reported  3.38-year 
periodic variations superimposed on the 1.4-GHz slow rise in the radio 
emission from \WRE. However, as the authors stated these variations 
are too short to be the WR+O binary period and might be caused by a 
third, low-mass, object in the system.

The optical extinction of A$_V = 8.32$~mag was derived
from detailed quantitative analysis of optical and infrared
emission from \WR \citep{dessart_00}, implying a foreground column
density of N$_H = (1.37-1.85)\times10^{22}$~cm$^{-2}$.
The range corresponds to the conversion that is used:
N$_H = (1.6-1.7)\times10^{21}$A$_{\mbox{V}}$~cm$^{-2}$
(\citealt{vuong_03}; \citealt{getman_05});
and 
N$_H = 2.22\times10^{21}$A$_{\mbox{V}}$~cm$^{-2}$
\citep{go_75}.
We adopt the stellar wind parameters (velocity and mass loss) of
the WR component 
V$_{WR} = 2700$\kms and 
$\dot{M}_{WR} = 3.15\times10^{-5}$\dotM ;
and of the O-star component
V$_{O} = 1500$\kms and  
$\dot{M}_{O} = 6.32\times10^{-6}$\dotM
that are based on the analysis by \citet{doug_00} and
\citet{dessart_00}.
From the same analysis, the distance to \WR is 1.4 kpc, which results
in projected (or minimum) binary separation of $226.8\pm11.2$ au
($0\farcs162\pm0\farcs008$).

In X-rays, a marginal detection was reported from the {\it Einstein}
survey of Wolf-Rayet stars, 
$6^{+5}_{-4}  \times10^{-3}$ cts s$^{-1}$ \citep{po_87}, 
and also from the \Rosat survey of Wolf-Rayet 
stars,  $3.5 \pm 1.9 \times10^{-3}$ cts s$^{-1}$ \citep{po_95}.
However,
both data sets had poor photon statistics ($\leq 30$ source counts).
\citet{rauw_15} reported a clear detection 
($\sim 1870$ source counts) of \WR in their
analysis of X-ray emission from massive stars in the star-forming
region Cygnus OB2. A two-temperature plasma 
(kT $=  0.36$ and $2.1$ keV)
could acceptably represent the observed \Chandra ACIS-I spectrum of
\WRE. By comparing with the {\it Einstein} and \Rosat data (mentioned
above), the authors also concluded that a long-term (years) 
variability might be present but the uncertainties in the old data 
prevent any firm conclusions on the variations of the X-ray flux from 
this object.

\section{Observations and data reduction}
\label{sec:observations}
We searched the \Chandra and \XMM archives for data on \WR and found
the following \Chandra ACIS-I data sets suitable for analysis of the 
X-ray emission from this object. Namely, one pointed observation (\WR
located on axis) taken on 2007 March 17 (ObsId: 7426) with effective 
exposure of 19.7 ks and two observations being part of the \Chandra 
Cygnus OB2 Survey (ObsId: 10967 and 10968; \WR located correspondingly 
at 9\farcm67 and 7\farcm86 off axis) both obtained on 2010 March 2 with 
the same effective exposure of 28.9 ks. We will further refer to these
observations as Obs 1, Obs 2 and Obs 3 (Fig.~\ref{fig:image}).

Following
the Science Threads for Imaging Spectroscopy in the 
CIAO 4.8\footnote{Chandra Interactive Analysis of Observations
(CIAO), \url{http://cxc.harvard.edu/ciao/}} data analysis software, we
extracted the \WR X-ray spectra from the three data sets (we
have initially re-processed the data adopting the CIAO 
{\sc chandra\_repro} script).
We note that the near-by source BD$+40 4243$ (11\farcs3 ~from \WRE) 
was also detected in Obs 1 (24 source counts). Although it is 
considerably fainter than \WRE,
we excluded its contribution to the X-ray 
spectra of \WR extracted from Obt 2 and Obs 3 (denoted by the red
circle in Fig.~\ref{fig:image}).
The \WR source counts were 719 (Obs 1), 773 (Obs 2) and 784 (Obs 3).
The \Chandra calibration database CALDB v.4.7.2 was used to
construct the response matrices and the ancillary response files.

We constructed the background-subtracted light curves and found no
variability on time scales shorter than the effective exposure of each
observation.
Namely, using time bins between 100 and 1000 s, the X-ray LCs were 
statistically consistent with a constant flux:
adopting $\chi^2$ fitting, the LCs were fitted with a constant and the 
goodness of fit was $\geq 0.94$ (Obs 1), $\geq 0.95$ (Obs 2) and 
$\geq 0.60$ (Obs 3).
Anticipating the results from our analysis, we note that
the observed flux from \WR was the same in Obs 1, Obs 2 and Obs 3.

For the spectral analysis in this study, we made use of 
standard as well as custom models in versions 11.3.2 and 12.9.1 of 
\xspec \citep{Arnaud96}.

\section{CSW model spectra}
\label{sec:csw_spec}

\subsection{CSW model}
\label{subsec:model}
In this study, we consider CSW picture that results from interaction
of two spherically symmetric gas flows (stellar winds) that have
reached their terminal velocities in front of the shocks. This is well
justified in wide binary systems (as \WRE), where neither radiative 
braking nor orbital motion (orbital velocities are much less than the
wind velocities) is expected to play an important role. In such a
case, the interaction region has cylindrical symmetry and
two-dimensional (2D) numerical hydrodynamic models are well suited for
calculating the physical parameters of the CSW structure.

We recall that the basic input parameters for the CSW hydrodynamic 
model in WR$+$O binaries are the mass loss and velocity of the stellar 
winds of the binary components and the binary separation
(\citealt{lm_90}; \citealt{luo_90}; \citealt{stevens_92};
\citealt{mzh_93}).
The former define a dimensionless parameter 
$\Lambda = (\dot{M}_{WR} V_{WR}) / (\dot{M}_{O} V_{O})$ which
determines the shape and the structure of the CSW interaction region 
\citep{mzh_93}.

Given the wind parameters and the binary separation
(Section~\ref{sec:wrstar}), we see that the shocked plasma in
the CSW region of \WR will be adiabatic.
It follows from the values of either of dimensionless parameters 
$\chi$ \citep{stevens_92} and $\Gamma_{ff}$ \citep{mzh_93}: 
$\chi = 574.0$ ($\chi > 1$ - adiabatic case),
$\Gamma_{ff} = 0.0002$ ($\Gamma_{ff} > 1$ - cooling is important).
In general, partial electron heating might occur behind strong shocks,
and the value of the dimensionless parameter $\Gamma_{eq} = 0.02$ 
indicates that this should be taken into account in the case 
of \WR ($\Gamma_{eq} < 1$ if the difference of electron and ion 
temperatures is important; see \citealt{zhsk_00}). 
Also, the value of the dimensionless parameter $\Gamma_{NEI} = 1.55$
indicates that the non-equilibrium ionization effects (NEI) could
play an important role in the CSW region of \WR as well (the NEI 
effects can be neglected if $\Gamma_{NEI} \gg 1$; see \citealt{zh_07}).

We therefore made use of our CSW \xspec models that take into account 
the different ion and electron temperature behind the shocks
\citep{zhsk_00} and the NEI effects in hot plasmas \citep{zh_07}.
These models are based on the 2D numerical hydrodynamic model of CSW
by \citet{lm_90} (see also \citealt{mzh_93}). The latter adopts the
`shock fitting' technique, which provides an exact solution to all 
discontinuity surfaces (the two shocks and the contact discontinuity)
of the CSW region.
This means that there is no numerical `mixing' of the shocked gases of
the stellar winds, which in turn allows the different chemical
composition of the WR and O-star wind to be explicitly taken into 
account in modelling the X-ray emission from the CSW region in \WRE.

We recall that the entire fitting procedure is threefold: 
(a) given the stellar wind and binary parameters, the hydrodynamic
model provides the physical parameters of the CSW region;
(b) based on these results, we prepare the input quantities 
(e.g., distribution of temperature, emission measure, ionization age) 
for the {\it spectral} model in \xspecE;
(c) the CSW \xspec  model fits the observed X-ray spectrum.
Note that in \xspec we can fit for the X-ray absorption, chemical
abundances and the model normalization parameter.
{\it 
If adjustments of other physical parameters are required, all three
steps should be repeated: i.e., our fitting procedure is an iterative
process.
It is important to keep in mind that the normalization parameter
($norm$) of the CSW model in \xspec is a dimensionless
quantity that gives the ratio of observed to theoretical fluxes. 
Thus, the entire fitting procedure is aimed at getting a value of 
$norm = 1.0$, which indicates a perfect match between the 
observed count rate and that predicted by the model 
($norm < 1.0$ indicates a theoretical flux higher than that required 
by the observations and the opposite is valid for $norm > 1.0$).
This in turn helps us obtain some constraints on the basic parameters 
of the CSW picture (e.g., mass-loss rates, binary separation).
}

Finally, we note that our CSW models were originally developed in
the \xspec version 11.3.2. The models were now `transferred' into
\xspec version 12.9.1. We tested the CSW models in both \xspec 
versions using the latest atomic data (as in \xspec 12.9.1) and the 
model results were identical.

\begin{figure*}
\begin{center}
\includegraphics[width=2.8in, height=2.0in]{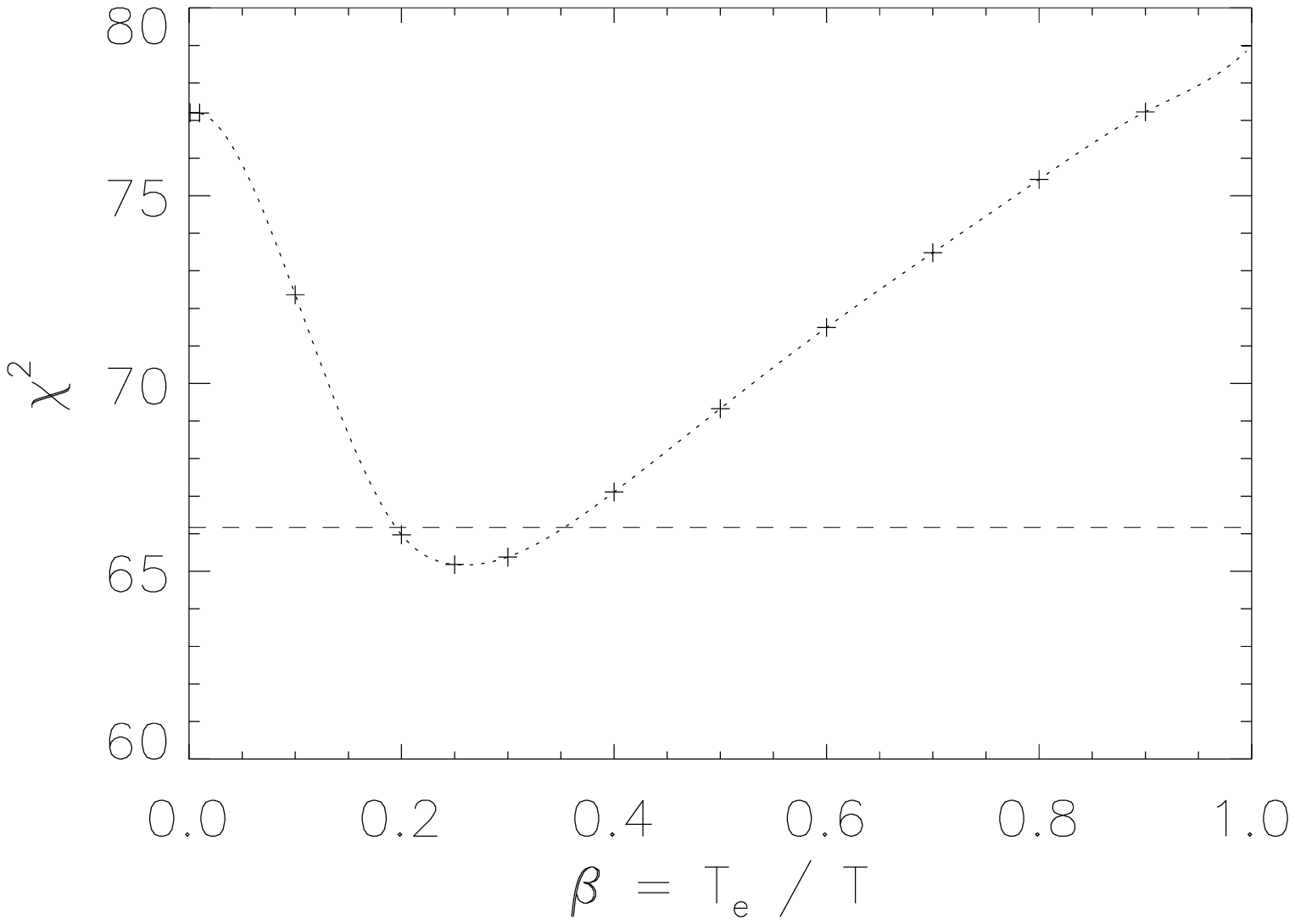}
\includegraphics[width=2.8in, height=2.0in]{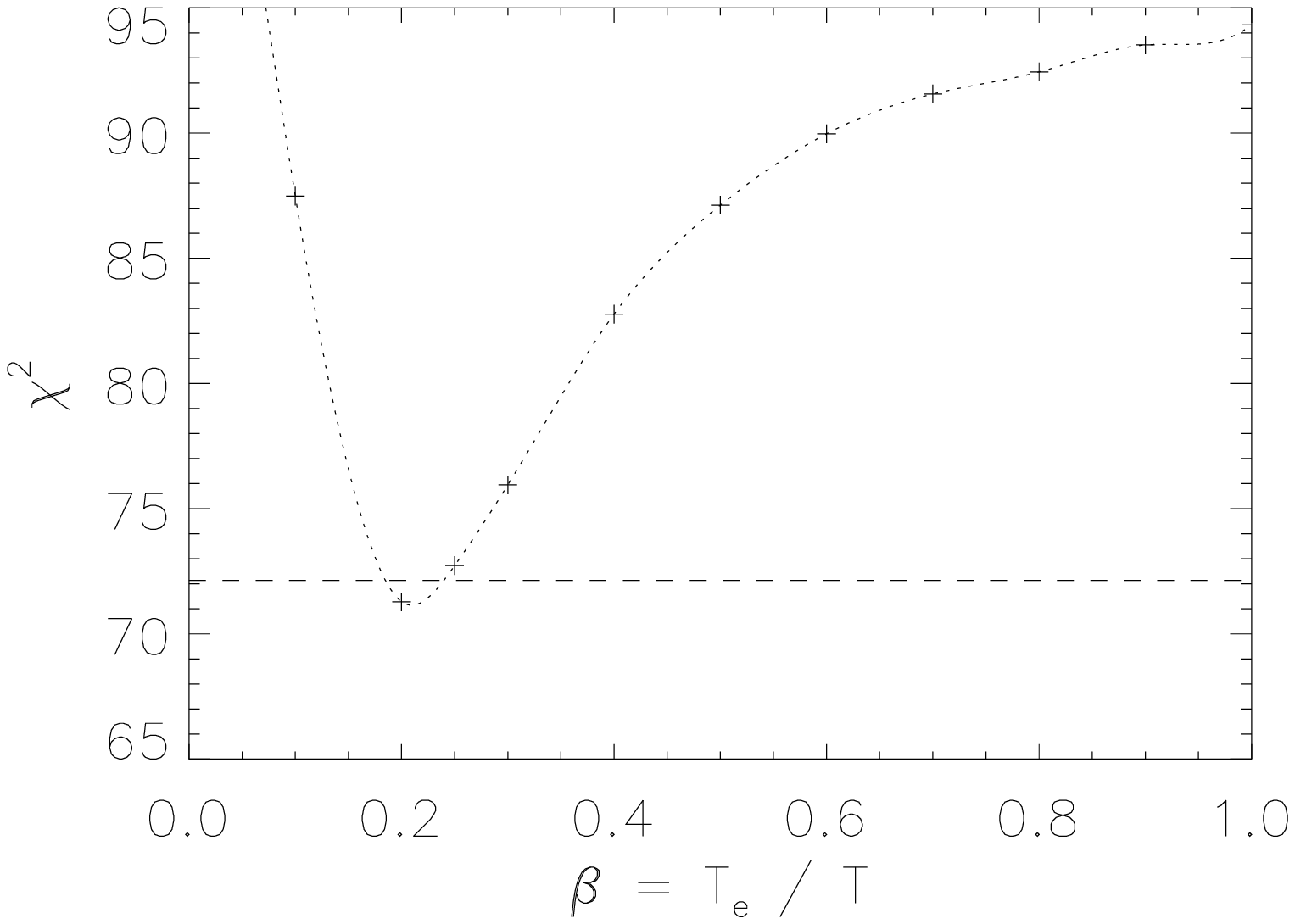}
\includegraphics[width=2.8in, height=2.0in]{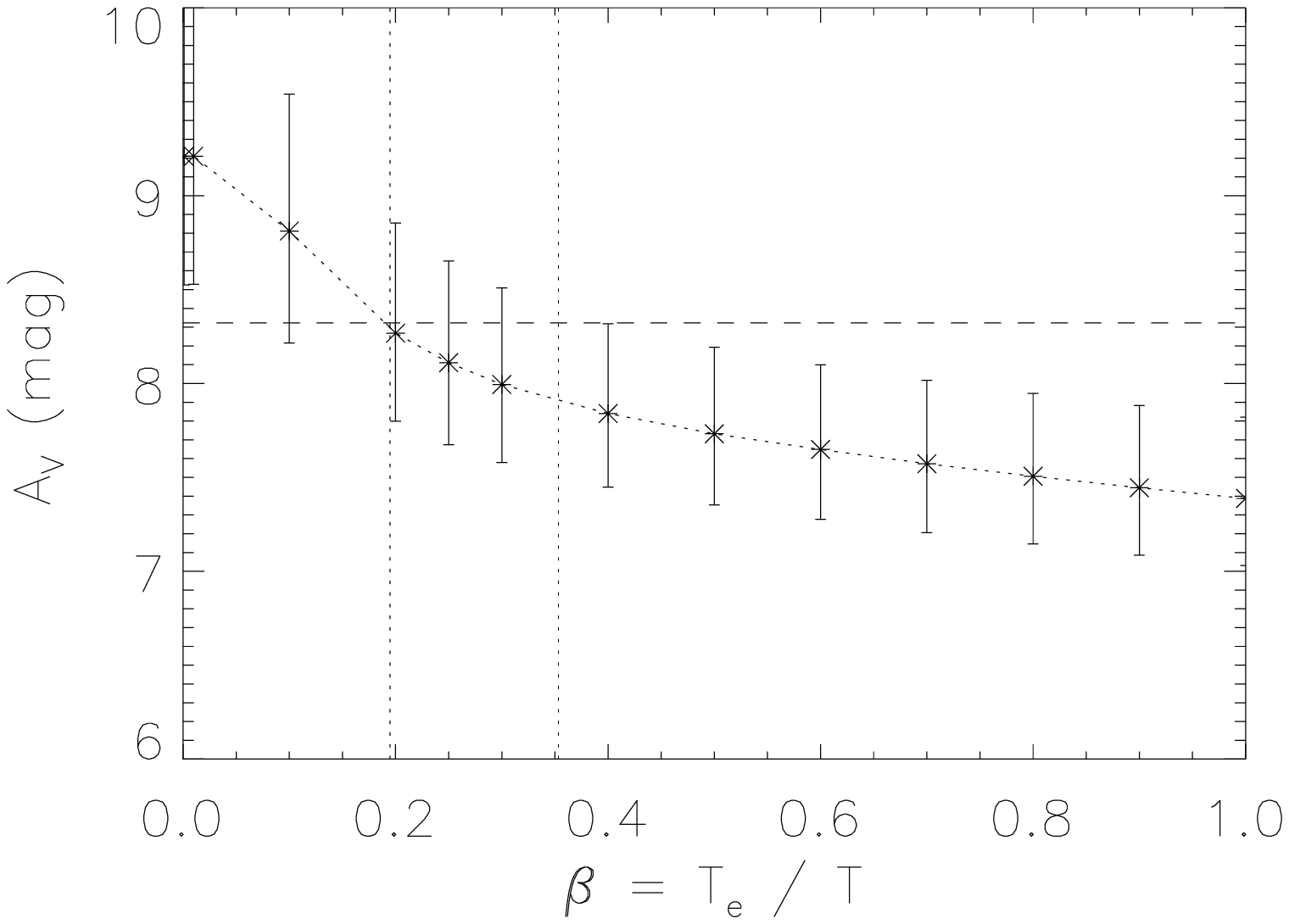}
\includegraphics[width=2.8in, height=2.0in]{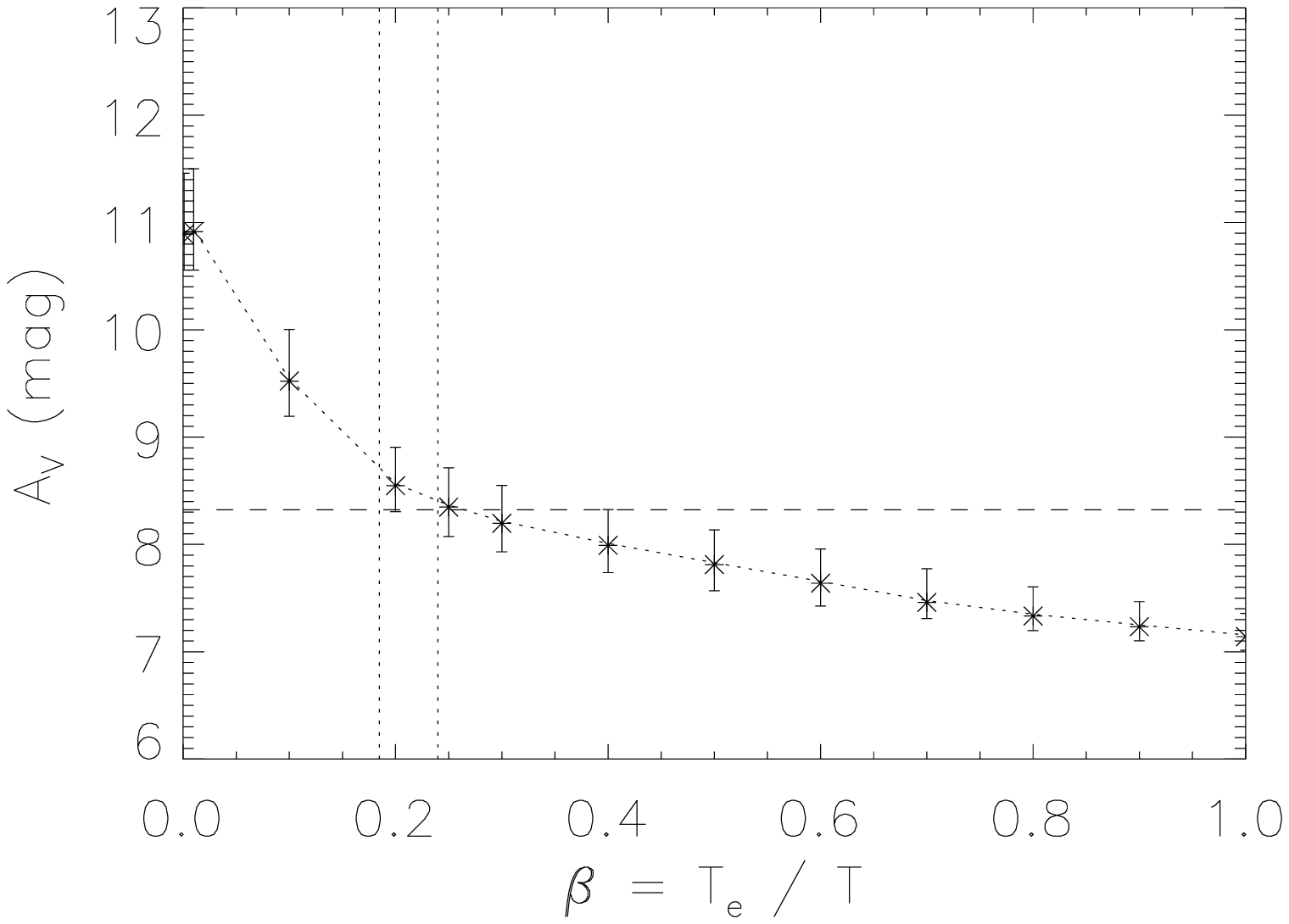}
\includegraphics[width=2.8in, height=2.0in]{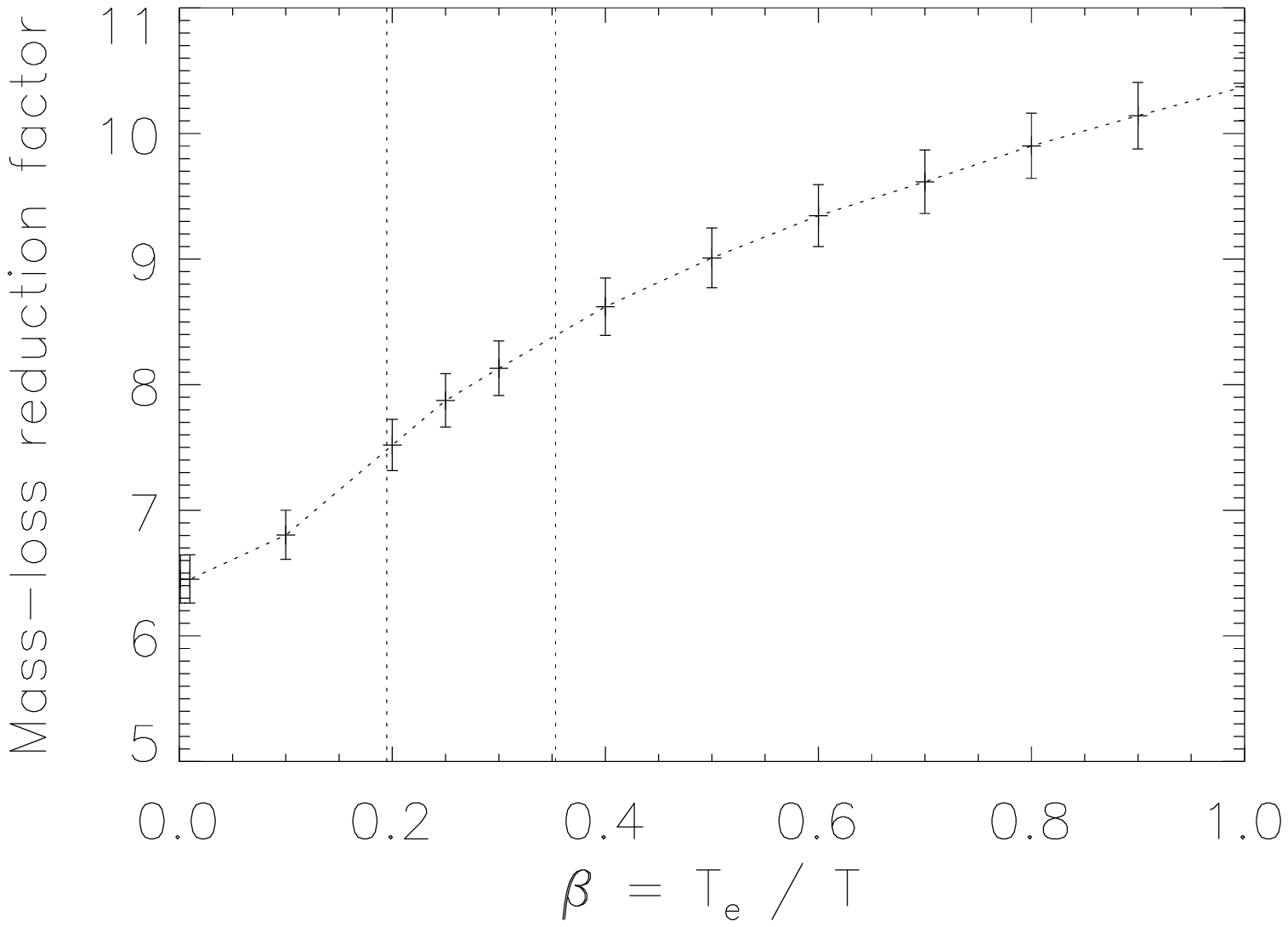}
\includegraphics[width=2.8in, height=2.0in]{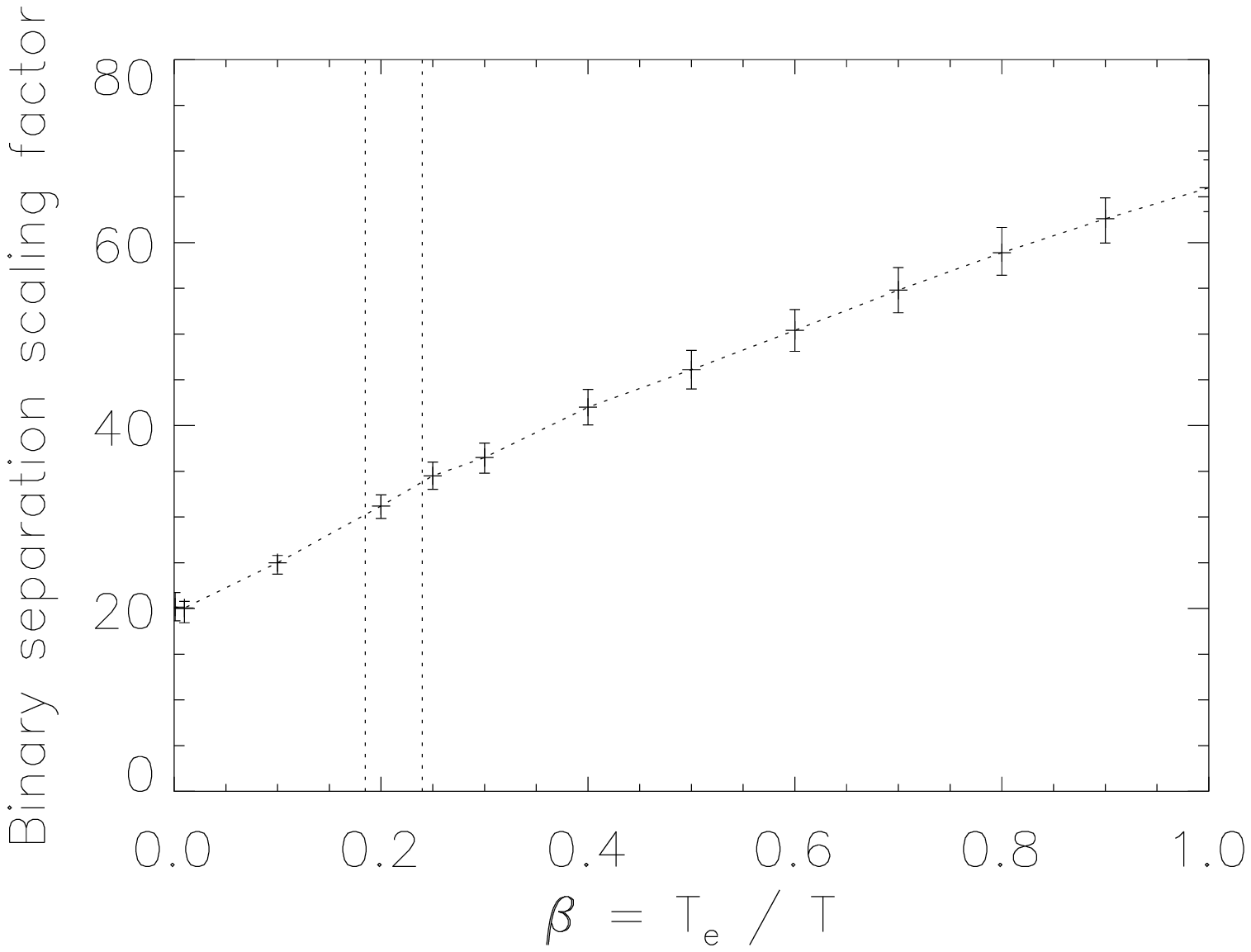}
\end{center}
\caption{Some parameters from the CSW model fits to the X-ray spectra
of \WR as function of the partial heating ($\beta = T_e / T$)  at 
the shock fronts. 
Left-hand panels: the case with reduced mass-loss rate.
Right-hand panels: the case with increased binary separation. 
The horizontal dashed line in the $\chi^2$-plots is at a value of 
$\chi^2_{min} + 1$, thus, illustrating the formal $1\sigma-$boundaries 
for the `best'-fit $\beta$ case. These boundaries are indicated by two
vertical dashed lines on the other plots.
The A$_V$ values in the plots are
calculated from the values of the X-ray absorption column densities
derived in the fits using the Gorenstein conversion \citep{go_75}.
The horizontal dashed line in the A$_V$-plots corresponds to the value
of optical extinction derived from analysis of optical and infrared
spectra of \WR \citep{dessart_00}. 
}
\label{fig:csw_results}
\end{figure*}

\begin{table*}
\caption{CSW Spectral Model Results 
\label{tab:csw}}
\begin{tabular}{llllll}
\hline
\multicolumn{1}{c}{Parameter} & 
\multicolumn{1}{l}{`Basic'} & 
\multicolumn{1}{c}{`$\dot{M}$'} &
\multicolumn{1}{c}{`D'} & 
\multicolumn{1}{l}{`$\dot{M}$' $ +$ T$_{ei}$} &
\multicolumn{1}{l}{`D' $+$ T$_{ei}$} 
\\
\multicolumn{1}{c}{} & \multicolumn{1}{c}{} &
\multicolumn{1}{c}{(A1)} & \multicolumn{1}{c}{(A2)} &
\multicolumn{1}{c}{(B1)} & \multicolumn{1}{c}{(B2)} \\
\hline
Reduced $\dot{M}$ by a factor of  &  1.0  &
    $10.4^{+0.3}_{-0.3}$  & 1.0  &  $7.9^{+0.2}_{-0.2}$ & 1.0 \\
Increased D by a factor of  & 1.0  & 1.0   & 
    $66.0^{+3,3}_{-2.6}$  & 1.0  &  $32.0^{+1.3}_{-1.3}$\\
$\beta = T_e / T$  &   1.0  &  1.0  &  1.0  &  
$0.26^{+0.10}_{-0.06}$ & $0.21^{+0.03}_{-0.03}$ \\
$\chi^2$/dof  &  85/103 & 79/103 & 94/103 & 65/103 &  71/103  \\
\vspace{0.03in}
N$_{H}$ (10$^{22}$ cm$^{-2}$)  &  
          1.23$^{+0.11}_{-0.08}$ &
          1.64$^{+0.10}_{-0.08}$ & 1.59$^{+0.04}_{-0.03}$ & 
          1.79$^{+0.12}_{-0.10}$ & 1.89$^{+0.08}_{-0.06}$ \\ 
\vspace{0.03in}
Ne  & 0.56$^{+0.50}_{-0.30}$
    & 0.35$^{+0.18}_{-0.13}$ & $\leq 0.06$  
    & 0.33$^{+0.23}_{-0.16}$ & 0.11$^{+0.11}_{-0.08}$  
    \\ 
\vspace{0.03in}
Mg  & 1.06$^{+0.32}_{-0.24}$
    & 0.48$^{+0.11}_{-0.10}$ & 0.17$^{+0.07}_{-0.08}$  
    & 0.29$^{+0.12}_{-0.10}$ & 0.10$^{+0.07}_{-0.06}$  
    \\ 
\vspace{0.03in}
Si  & 2.15$^{+0.60}_{-0.53}$
    & 0.86$^{+0.24}_{-0.22}$ & 0.11$^{+0.24}_{-0.11}$  
    & 0.41$^{+0.25}_{-0.23}$ & $\leq 0.09$  
    \\
\vspace{0.03in}
$norm$ & 0.0073$^{+0.0004}_{-0.0004}$
       & 1.00$^{+0.05}_{-0.05}$ & 0.99$^{+0.05}_{-0.04}$ 
       & 1.00$^{+0.05}_{-0.05}$ & 1.00$^{+0.04}_{-0.04}$  
    \\
F$_{X}$ ($10^{-13}$ ergs cm$^{-2}$ s$^{-1}$)  &
           \,\,4.14  &
           \,\,4.16  & \,\,4.25  &
           \,\,3.69  & \,\,3.62  \\
                                              &
            (\,29.7) &
            (161.6) &  (390.4) &
            (233.7) &  (675.9) \\
\hline

\end{tabular}

Note --
Results from  simultaneous fits to the \Chandra
spectra of \WR using model spectra from the CSW hydrodynamic 
simulations. 
The model with nominal stellar wind and binary parameters is denoted
`Basic'.
The models with reduced mass-loss rates are denoted with `$\dot{M}$'
and
those with increased binary separations are denoted with `D'.
The models denoted with $ +$ T$_{ei}$ are their corresponding 
best-fit versions
that take into account the different electron and ion temperatures
(parameter $\beta$ gives the partial heating at the shock front).
For each model, given is the factor by which 
the mass-loss rates of the stellar winds were reduced or the binary
separation was increased (see Section 
\ref{subsec:results} for details).
Tabulated quantities are the neutral hydrogen absorption column
density (N$_{H}$), the  Ne, Mg, and Si abundances, the 
normalization parameter ($norm$) and the absorbed X-ray flux (F$_X$) 
in the 0.5 - 10 keV range followed in parentheses by the unabsorbed 
value. The $norm$ parameter is a dimensionless
quantity that gives the ratio of observed to theoretical fluxes. A
value of $norm = 1.0$ indicates a perfect match between the observed
count rate and that predicted by the model.
The derived abundances are with respect to the typical \WR abundances 
(see text; Section~\ref{subsec:results}). 
Errors are the $1\sigma$ values from the \xspec fits
as those of the reduced $\dot{M}$ and increased $D$ factors have been
propagated from the $1\sigma$ values on the $norm$ parameter.
The errors on the $\beta$ parameter for the best-fit cases (models
B1 and B2) are from the entire fitting procedure (see text for
details).

\end{table*}

\subsection{CSW spectral model fits}
\label{subsec:results}
As an initial step, we fitted each of the spectra separately using
standard thin-plasma and shock models in \xspecE. We found that the 
observed X-ray flux was the same in all of them:
e.g., two-temperature fits provide the following observed fluxes in
the (0.5 - 10 keV) energy range 
$4.10 [3.35-4.16] \times10^{-13}$ ergs cm$^{-2}$ s$^{-1}$ (Obs 1),
$3.87 [3.16-3.99] \times10^{-13}$ ergs cm$^{-2}$ s$^{-1}$ (Obs 2),
$4.04 [3.27-4.16] \times10^{-13}$ ergs cm$^{-2}$ s$^{-1}$ (Obs 3)
as the $1\sigma$ confidence range is given in the square brackets.
That is there was no
long-term variability present. So to take advantage of the available 
photon statistics, we further fitted these 
spectra simultaneously by imposing the same CSW model parameters for
each individual spectrum.

In all the spectral fits, the chemical abundances of the shocked 
O-star wind were solar \citep{ag_89}. For the shocked WR-star wind, we 
adopted  the carbon and oxygen abundances from \citet{dessart_00}:  
C / He = 0.08; O / He = 0.02 by number; while the other elements had
their values typical for the WC stars \citep{vdh_86}. The Ne, Mg and 
Si abundances of the shocked WR plasma were allowed to vary to improve 
the quality of the fits.

Our basic CSW model was the one that adopts the nominal values of the
wind and binary parameters of \WR (see Section~\ref{sec:wrstar}).
The theoretical spectrum matched well the shape of the 
observed spectra but it overestimated the observed flux considerably
which means that the observations require much smaller emission 
measure in the CSW region than that the nominal \WR parameters 
suggest.

We recall that the emission measure in the CSW region that results
from interaction of spherically symmetric stellar winds is 
proportional to the square of the stellar wind mass loss and 
is reversely proportional to the binary separation
(EM $\propto n^2 V$, $n$ is the number density, $V$ is
the volume; $n \propto \dot{M}/a^2$ and $V \propto a^3$,
therefore EM $\propto \dot{M}^2/a$).
We thus iterated through our fitting procedure, that is we
repeated all three steps ot its numerous times by adopting
either reduced mass-loss rates or increased binary separation 
until the CSW spectral models provided
excellent correspondence between theoretical and observed fluxes
for each individual case
(i.e., having normalization parameter of the \xspec model $norm
\approx 1$).
Similarly, we explored a range of values for the partial heating of the
electrons at the shock fronts ($\beta = T_e / T$, $T_e$ is the
electron temperature and $T$ is the mean plasma temperature), since 
both decrease of mass loss and increase of binary separation make the 
electron-ion temperature equalization run slower downstream from the 
shock front due to the reduced plasma density (e.g., see eq.1 in 
\citealt{zhsk_00}).

\begin{figure*}
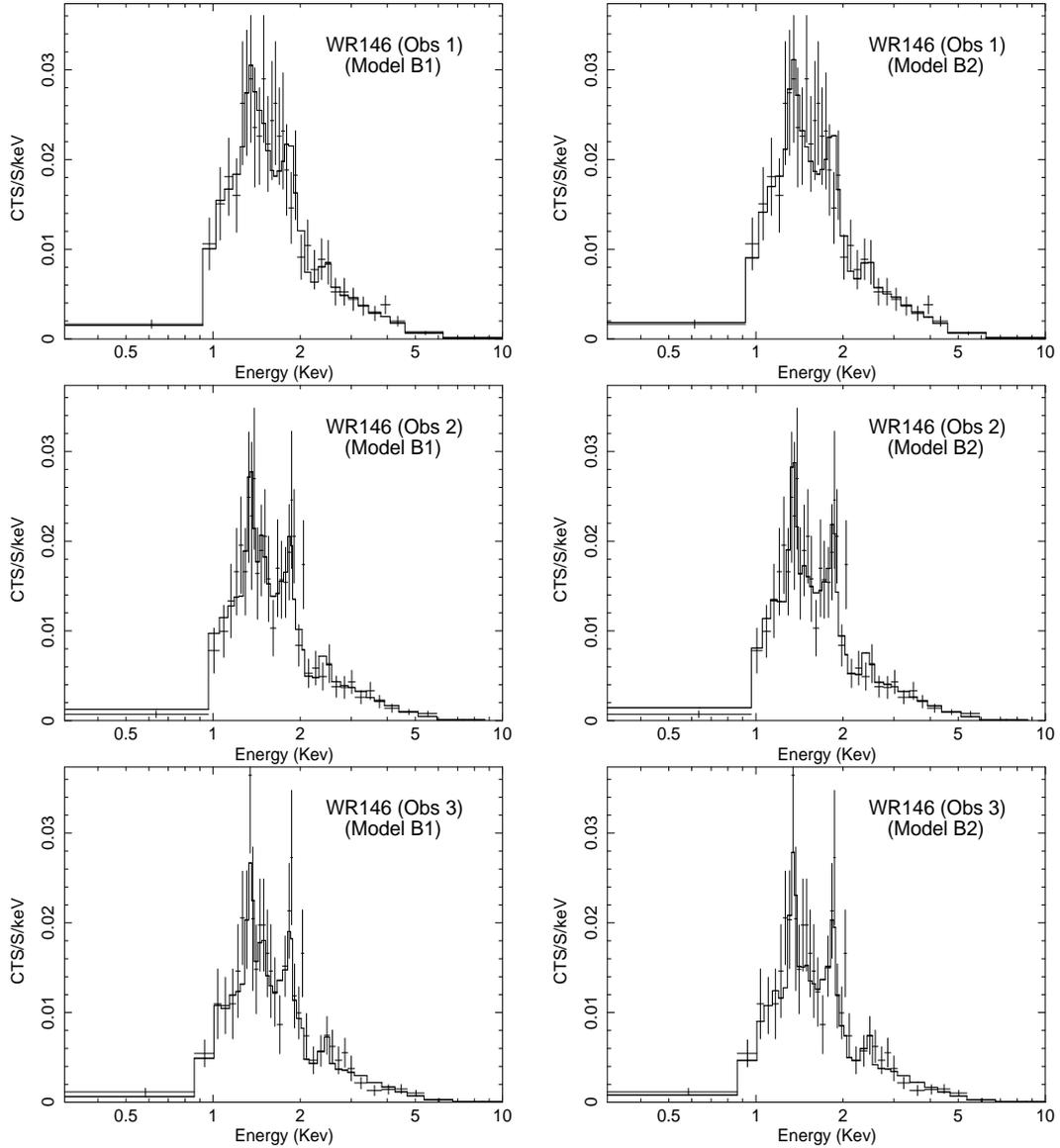

\begin{center}
\includegraphics[width=2.0in, height=2.8in,angle=-90]{fig3a.eps}
\includegraphics[width=2.0in, height=2.8in,angle=-90]{fig3d.eps}
\includegraphics[width=2.0in, height=2.8in,angle=-90]{fig3b.eps}
\includegraphics[width=2.0in, height=2.8in,angle=-90]{fig3e.eps}
\includegraphics[width=2.0in, height=2.8in,angle=-90]{fig3c.eps}
\includegraphics[width=2.0in, height=2.8in,angle=-90]{fig3f.eps}
\end{center}
\caption{
The background-subtracted spectra of \WR  overlaid with the
CSW model `best' fits.  
The case with reduced mass-loss rate is shown in the left-hand panels 
while that with increased binary separation in the right-hand panels: 
respectively models (B1) and (B2) from Table~\ref{tab:csw}.
The spectra were re-binned to have a minimum of 20 counts per bin.
}
\label{fig:spec_csw}
\end{figure*}

Some fit results are summarized in Figure~\ref{fig:csw_results}, while
the `best'-fit results are given in Table~\ref{tab:csw} and
Figure~\ref{fig:spec_csw}.

Although all the models are acceptable in formal statistical sense
(see both top panels in Fig.~\ref{fig:csw_results}),
we note that the `smooth' $\chi^2$ curve illustrates the similarity of
our fitting procedure and the {\sc steppar} command in 
\xspecE\footnote{
\url{https://heasarc.gsfc.nasa.gov/docs/software/lheasoft/xanadu/xspec/manual/node87.html}}.
Namely, we perform a fit while stepping the value of a parameter 
($\beta$, the partial electron heating in this case) in a given range. 
This allows us to derive the best-fit value  (determined
by the minimum of the $\chi^2$ statistic, $\chi^2_{min}$) and the
corresponding confidence range for this parameter.
For example, the formal $1\sigma$ errors come in the
usual way, namely: $\chi^2_{1\sigma} = \chi^2_{min} + 1 $. 
We see from Fig.~\ref{fig:csw_results} that the CSW models with partial 
electron heating at the shock
fronts ($\beta < 1$)  provide better fits to the observed X-ray
spectra of \WRE. The best fits are with 
$\beta = 0.26^{+0.10}_{-0.06}$
and 
$\beta = 0.21^{+0.03}_{-0.03}$
for the model series with reduced mass loss and increased binary 
separation, respectively
(denoted by the
dashed lines in the right-hand side panels in
Fig,~\ref{fig:csw_results}). 
For these best fits, the  mass-loss rates are correspondingly
reduced by a factor of $7.9^{+0.5}_{-0.4}$ and the binary 
separation is correspondingly increased by a factor of 
$32.0^{+2.0}_{-1.8}$ with respect to their nominal values (see
Section~\ref{sec:wrstar}).
We note that thus derived uncertainties on the mass-loss rate and 
binary separation factors are larger than those  from individual 
fits since the latter reflect only the quality of data.

On the other hand, we think that the formal errors on the derived
$\beta$-values are not that important while it is in fact important
that CSW models with partial electron heating are a better
representation of the observed X-ray spectrum of the CSW binary \WRE.
Interestingly, similar conclusion is valid for other {\it wide} CSW 
binaries as WR~140 (see Table~3 in \citealt{zhsk_00}), WR 137
(see Table~2 in \citealt{zh_15}) 
and CygOB2 9 \citep{parkin_14}.

Also, the derived values of the X-ray absorption for the
`best'-fitting
models are consistent with the optical extinction deduced from
analysis of the optical and infrared emission of \WR
\citep{dessart_00}. Thus, no excess absorption, that might be due to
the massive stellar winds, is indicated from analysis of the X-ray 
emission from this object in the framework of the CSW 
picture\footnote{
Note that our X-ray analysis may indicate some excess absorption if 
the \citet{vuong_03} and \citet{getman_05} conversions were adopted 
(see Section \ref{sec:wrstar}). However, these conversions are based 
on analysis {\it only} of nearby dense clouds ($d < 500$ pc) and they 
deviate from the galactic conversion \citep{vuong_03}. We thus give
preference to the \citet{go_75} conversion.
}.

But, the most interesting result from the direct confrontation of the 
CSW models and the X-ray spectra of \WR is probably that considerably
reduced mass-loss rates (by a factor of $\sim 8$) and/or 
increased binary separation (by a factor of $\sim 30$) are
required to have a good correspondence between the CSW theory and
observations.
This result deserves some more discussion, we believe, thus,
we will return to it in Section~\ref{sec:discuss}. 
However, we would also like to explore the variability issue a bit 
more before continuing our discussion.

\begin{figure}
\begin{center}
\includegraphics[width=2.0in, height=2.8in,angle=-90]{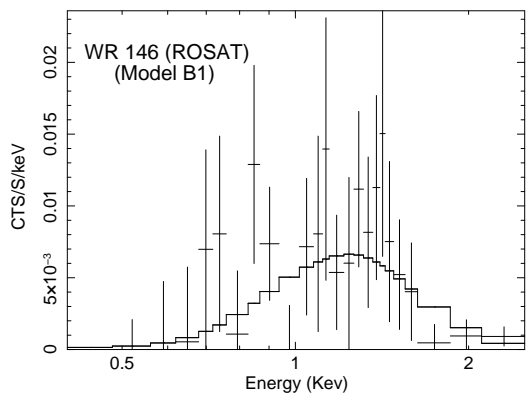}
\end{center}
\caption{
The background-subtracted \Rosat spectrum of \WR  overlaid with the
CSW model B1 that gives excellent fit to the \Chandra spectra of this
CSW binary (see Table~\ref{tab:csw}).
The spectrum was re-binned to have a minimum of 10 counts per bin.
}
\label{fig:spec_ros}
\end{figure}

\subsection{On the long-term X-ray variability of \WR}
\label{subsection:rosat}
We note that the lack of relatively long-term
variability of its X-ray emission (Section~\ref{sec:wrstar} and
~\ref{subsec:results}) is something to expect in the case of CSW wide
binaries, unless the object has highly eccentric orbit with not very 
long binary period (e.g., a few years) as is the case of the prototype
CSW binary, the WR+O system WR~140 (\citealt{williams_90};
\citealt{williams_11} and the references therin).
\citet{rauw_15} reported that there is no appreciable sign of
variability between the {\it Einstein} and \Chandra observations of
\WRE, but the case of \Rosat observations was not completely settled.

We searched the \Rosat archive and found that \WR fell in the \Rosat
field of view on 1993 April 29: PSPC data set rp900314n00
(30\farcm6  off axis) with nominal exposure of 19.4 ks.
Following the recommendations for the \Rosat Data
Processing 
\footnote{\url{https://heasarc.gsfc.nasa.gov/docs/rosat/rhp_proc_analysis.html}}, 
we extracted the source and background spectra. The net source counts
were $121\pm20$ in an effective exposure of 18.7 ks.
Since the data were taken after 1991 Oct 14, we adopted the response 
matrix pspcb\_gain2\_256.rmf and we
used the package {\sc pcarf} to construct the corresponding ancillary 
response file. 

Although the quality of the spectrum is not high it is much better
than that of the data used in the \Rosat survey of WR stars (see the 
comment on the X-ray counts in Section~\ref{sec:wrstar}), so, we 
decided to use this \Rosat spectrum to check the long-term variability 
of \WRE. Figure~\ref{fig:spec_ros} shows the \Rosat spectrum overlaid 
with one of the CSW models (model A2 in Table~\ref{tab:csw}) that 
matches very well the \Chandra spectra of this object. We see a good 
correspondence between the CSW model and \Rosat observation: 
$\chi^2 / dof = 17 / 25$; observed count rate of 
$6.5\times10^{-3}$ cts s$^{-1}$ vs. model
count rate of $6.0\times10^{-3}$ cts s$^{-1}$.
Also, if we kept the spectral shape unchanged (i.e. abundances and 
X-ray absorption fixed) and varied only the normalization parameter of 
the model, the best-fitting \Rosat flux (`required' CSW emission
measure) was within 4 per cent of its
nominal (as of model A2) value. This, in conjunction with the results 
by \citet{rauw_15},  makes us confident to conclude that
there is no indication for a long-term (of approximately two-three 
decades) variability in the X-ray emission from \WRE.
We note that this corresponds well to the {\it CSW wide binary} 
status of this object.
Namely, from the Kepler's third law and assumed reasonable total mass 
of the binary components of 20-40 solar masses, the projected 
(minimum) binary separation of 226.8 au (Section~\ref{sec:wrstar}) 
suggests a binary period of more than $\sim 550$ years. Thus, the 
available X-ray observations 
(from {\it Einstein}-through-\RosatE-to-\ChandraE) cover
less than 5 per cent of that, and no considerable changes of the
binary separation and local X-ray absorption (if any, see
Section~\ref{subsec:results}) are expected over such a small piece of
the orbit in a {\it wide} binary system.

\section{Discussion}
\label{sec:discuss}
\WR is a {\it wide} CSW binary 
(Section~\ref{sec:wrstar}), thus, the goal of this study was to carry
out {\it direct} comparison between the CSW model spectra and the
X-ray observations. 
It is worth noting that we have attributed the entire X-ray emission
from \WR to some X-ray production mechanism that may operate in 
massive binaries: the CSW phenomenon in this case. This is a
reasonable assumption, we believe, since the WR star in this system is
a carbon-rich object and the WC stars are very faint or X-ray quiet
objects: all the pointed observations of presumably single WC stars
resulted in non-detections (\citealt{os_03}; \citealt{sk_06}).
On the other hand, massive OB stars are X-ray sources and we could
expect no more than a few $\times 10^{32}$ ergs s$^{-1}$ from the
O-star companion in \WRE, provided it had a bolometric luminosity even
as high as $10^{39} - 10^{40}$ ergs s$^{-1}$:
\citet{rauw_15} have shown that the OB stars in CygOB2 (\WR being a
probable member of this star-forming region) follow the
X-ray-to-bolometric luminosity relation 
$\log (L_X / L_{bol}) = -7.2\pm0.2$.
Such a possible contribution 
is not substantial compared to that from the CSW region in \WR
($L_X > 3.8\times10^{33}$ ergs s$^{-1}$ for an adopted distance of 1.4
kpc, using the unabsorbed fluxes from
Table~\ref{tab:csw})
but even if it were, then what about the X-ray 
emission from the CSWs themselves? 

The very basic result from the CSW model analysis is the `requirement' 
of considerable changes of the stellar mass-loss rates and/or binary
separation with respect to their currently accepted values to make the 
theory and observations converge.

In fact, we considered in some detail two limiting
cases that gave very good correspondence between the theoretical and
observed flux (emission measure): (a) reduced mass losses; (b) 
increased binary separation (Section~\ref{subsec:results}). However, 
it is our understanding that case (b) does not seem very realistic 
since it could suggest a very `special' (in general possible but with
low probability)  observational circumstances: e.g., the orbital 
inclination angle of the spatially-resolved binary \WR should be very 
close to $90^\circ$ (only within $2^\circ$ of it), for the
expected (i.e. actual) 
binary separation is at least as  $\sim 30$ times as large 
(Table~\ref{tab:csw} and Fig.~\ref{fig:csw_results}) the measured 
(projected) one.

In general, other combinations of mass-loss rates and orbital
separation with values in the range we have considered separately for 
each of them that give a reduction of the emission measure similar to
that used in our analysis (EM $\propto \dot{M}^2/a$) can give 
acceptable fits to the X-ray spectra of \WRE. However, a reasonable 
value of the binary separation (say, $\sim 2 - 3$ times its minimum 
value) will still require mass-loss rates considerably reduced by a 
factor of 4 - 5 (or even more) compared to their currently accepted 
values for this CSW binary. {\it But, is \WR unique in this respect?}

We recall that one of the basic results of the {\it direct} modelling 
of the X-ray emission from the {\it wide}  CSW binaries WR~137 and 
WR~147 in the framework of the CSW picture was along the same lines. 
Namely, an appreciable reduction of the mass-loss rates was needed to 
reconcile the model predictions with observations: of about one order 
of magnitude for WR~137 \citep{zh_15}; and of about a factor of four
for WR~147 \citep{zhp_10b}. Thus, there is already some `statistics'
which gives indications for an appreciable mismatch between the 
stellar wind parameters of Wolf-Rayet stars derived from analysis of 
their optical/UV and radio emission and their corresponding values 
that are needed to explain the observed X-ray emission from the same 
objects. We have to keep in mind that the stellar wind parameters
(e.g., mass-loss rates) of
massive stars are in a way model-dependent since they are deduced by 
adopting some physical picture with related assumptions, 
approximations etc. 

The CSW picture in {\it wide} massive binaries is quite simple from a
technical point of view: it adopts adiabatic shocks that result from
interaction of homogeneous gas flows (stellar winds). On the other
hand, winds of massive WR stars are likley not smooth but clumpy and
this is adopted in the sophisticated stellar atmosphere models used to
derive their stellar wind properties (e.g., \citealt{hamann_06};
\citealt{sander_12}). 
In these models, a standard volume filling factor of 0.1 throughout
the stellar wind is usually assumed that results in a factor of $\sim
3 = 1/\sqrt{0.1}$ mass-loss reduction compared to models with
homogenious winds. We note that the mass-loss rate of \WR adopted in
this study (Section~\ref{sec:wrstar}) is from \citet{dessart_00}:
{\it 
it is based on spectral analysis with stellar atmosphere model with 
stellar-wind clumping taken into account. 
}

Using thus derived mass-loss rates in a CSW model that assumes
homogenious gas flows (stellar winds) means that we have explored a
physical picture in which the dense clumps expand further out from the
massive star itself. Thus, they `murge' with each other 
and form a homogenious stellar wind well beyond the UV/optical line
formation region but before the stellar winds collide in the wide 
binary system. 
However, even in such a case the adopted `basic' mass-loss rates of 
the stellar components in \WR produce way too high an X-ray emission 
from the CSW region compared to what is required by the observations.

Then, could it be that the stellar winds of WR
stars are two-component flows: the more massive component (dense
clumps) are responsible for the optical/UV emission from these objects
and the smooth rarefied component is a basic factor for their X-ray
emission that comes from the CSW region in WR binaries?

Although such a physical picture may seem speculative, it could in
general explain the discrepancy between the `optical/UV' stellar wind
parameters and their `X-ray' values required by the corresponding CSW
picture in {\it wide} binaries.
However, based on numerical hydrodynamic simulations of `clumpy' CSW 
in wide binaries  \citet{pittard_07} concluded that clumps will dissolve 
in the CSW region and the X-rays emission will be very similar to that
from smooth winds. Thus, introducing a two-component wind might not
resolve the mass-loss issue discussed here.

But, a similar mass-loss mismatch has also emerged from analysis of
the X-ray emission from wind-blown bubbles (WBB) around WR stars, that 
is from a different physical picture. Namely, theoretical hydrodynamic 
WBB models match very well the shape of the observed X-ray spectra but
they again `require' an appreciable reduction of the currently
accepted mass-loss rates for the central star of the studied WBB:
by a factor of $\sim 3-4$ for NGC~6888 \citep{zhp_11}; and almost by 
an order of magnitude for NGC~2359 \citep{zh_14}.

Summarizing this kind of results from modelling the X-ray emission
from wide CSW binaries and wind-blown bubbles, we could say that
they all seem to point to relatively small mass-loss rates from 
Wolf-Rayet stars, $\dot{M} < 10^{-5}$\dotM, and such values are 
in general atypical for these massive stars (e.g., see 
\citealt{crowther_07} for a review on the physical properties of 
WR stars).

We thus think that more efforts are needed to build a self-consistent
picture of the stellar winds from Wolf-Rayet stars that is based on
global modelling of the WR emission in different spectral domains: 
e.g., radio, optical/UV, X-ray.
Such an approach may also help reduce the uncertainties of the
distances to WR stars. On the one hand, these uncertainties are not
negligible since the distances to WR stars are not well constrained:
e.g., as a rule most distances to WRs are `photometric' \citep{vdh_01}.
And, on the other hand, these uncertainties directly affect the results
on the stellar wind parameters.

Along the same lines, could it be that such a mass-loss reduction is
required not only for the WR stars but for other massive stars of
early spectral type? For example, \citet{parkin_14} reported that 
the mass-loss rates of the binary components in the O$+$O binary Cyg
OB2 9 (Schulte 9) should be reduced by a factor of $\sim 7.5-7.7$
compared to their currently accepted values. This result is based on
analysis (hydrodynamic modelling) of the X-ray emission in the 
framework of the CSW picture. We would like to emphasize that the CSW 
shocks in Cyg OB2 9 are in {\it adiabatic} regime, as they are in the 
case of \WR and other CSW binaries with WR components discussed above. 
This is an important detail since the physics of adiabatic CSWs (thus, 
their corresponding numerical modelling) is relatively simple from a 
technical point of view which makes the corresponding numerical 
results quite reliable.

We note that if future studies {\it do} justify the need of a 
considerable (of about one order of magnitude) reduction of the 
mass-loss rates in massive stars of early spectral type, this may have 
an important impact on our understanding of the physics of these 
objects. Namely, we will have a new and deeper insight of the driving 
mechanism of their stellar winds and the evolution of these massive 
stars as well, since the latter considerably depends on how they lose 
mass during their life time. Also, the evolution of massive stars of 
early spectral type is an important ingredient for understanding the 
physics of young stellar clusters and star-forming regions where these 
stars are born and where they have huge impact on their environment.

However, caution is advised before making such an important turn. 
We have to see first if such a mass-loss discrepancy emerges in other 
objects, thoroughly evaluate all these basic findings, and only then 
proceed with so general conclusions. 

We thus think that a good approach for handling this issue is to 
adopt the global spectral modelling of the observational properties 
of massive stars (especially of Wolf-Rayet stars) in different 
spectral domains (e.g., radio, optical/UV, X-ray) as proposed above. 
This will show whether we could reconcile the results on mass-loss 
rates from different analyses, that is to see if a solution is 
possible all these results may converge on. Alternatively, such an 
approach may reveal some additional caveats in our understanding of 
the physics of stellar winds in massive stars or of the physics of 
fast shock, that result from the interaction of the highly supersonic flows 
(stellar winds) in massive binaries. Could it be that we are missing
some key detail of the CSW physics in massive binaries that has
a very important impact on the origin of X-rays in these objects?
We believe that the global spectral modelling will help us clarify all
that.

\section{Conclusions}
\label{sec:conclusions}
In this work, we presented an analysis of the X-ray emission from the
massive binary \WR {\it in the framework of the colliding stellar wind
picture}. The main results and conclusions from a {\it direct} 
comparison between the results from the numerical hydrodynamic CSW 
model and the \Chandra and \Rosat data on this object are as follows.

(i) 
We confirm (see also \citealt{rauw_15}) that
there are no indications of long-term (of approximately two 
decades) variability in the X-ray emission from \WRE. This is in 
accord with the CSW picture in {\it wide} (in fact, spatially 
resolved) massive binaries.

(ii) CSW model spectra match well the shape of the observed X-ray
spectrum of \WRE. 
There are indications that
models with partial electron heating at the
shock fronts (different electron and ion temperatures) with
$\beta = T_e / T = 0.2 - 0.3$ ($T_e$ is the
electron temperature and $T$ is the mean plasma temperature)
to be a better 
representation of the X-ray data than those with complete temperature 
equalization. 
Also, the derived X-ray absorption from the spectral
fits is consistent with the optical extinction to \WRE.

(iii) On the other hand, CSW models overestimate the observed X-ray
flux (emission measure) considerably. To reconcile the model
predictions and observations, the mass-loss rate of \WR must be
reduced by a factor of 8 - 10 compared to the currently accepted value
for this object (the latter already takes clumping into account).

(iv) Finally, the considerable mismatch between the mass-loss rates
based on X-ray studies and those from analysis in other spectral
domains for this and other WR stars makes us believe that we need to
build a self-consistent picture of the stellar winds from such massive 
stars adopting a {\it global spectral modelling} of their properties.

\section*{Acknowledgements}
This research has made use of data and/or software provided by the
High Energy Astrophysics Science Archive Research Center (HEASARC),
which is a service of the Astrophysics Science Division at NASA/GSFC
and the High Energy Astrophysics Division of the Smithsonian
Astrophysical Observatory. 
This research has made use of the NASA's Astrophysics Data System, and
the SIMBAD astronomical data base, operated by CDS at Strasbourg,
France.
S.A.Z. acknowledges financial support from Bulgarian National Science 
Fund grant DH 08 12.
The author thanks an anonymous referee for 
valuable comments and suggestions.

\bibliographystyle{mnras}
\bibliography{wr146}

\bsp    
\label{lastpage}
\end{document}